\documentclass[11pt,a4paper,portrait,times]{article}
\usepackage[numbers]{natbib}
\usepackage{amsmath}
\usepackage{eqnarray}
\usepackage{amssymb}
\usepackage[pdftex]{graphicx} 
\usepackage{float}
\usepackage{url} 
\usepackage{pdflscape}
\usepackage[retainorgcmds]{IEEEtrantools}
\usepackage{soul}

\title{ULF wave transmission across collisionless shocks: 2.5D local hybrid simulations}
\date{January 2022}

\begin{document}
\maketitle

\begin{center}
P. Kajdi\v{c}$^1$, Y. Pfau-Kempf$^2$, L. Turc$^2$, A. P. Dimmock$^3$, M. Palmroth$^2$, K. Takahashi$^4$, E. Kilpua$^2$, J. Soucek$^5$, N. Takahashi$^6$, L. Preisser$^1$, X. Blanco-Cano$^1$, D. Trotta$^7$, D. Burgess$^8$\\
{\small  $^1$Departamento de Ciencias Espaciales, Instituto de Geof\' isica, Universidad Nacional Aut\'onoma de M\'exico, Ciudad Universitaria, Ciudad de M\'exico, Mexico\\
$^2$Department of Physics, University of Helsinki, Helsinki, Finland\\
$^3$Swedish Institute of Space Physics (IRF), Uppsala, Sweden\\
$^4$The Johns Hopkins University Applied Physics Laboratory, Laurel, Maryland, USA\\
$^5$Institute of Atmospheric Physics, Academy of Sciences of the Czech Republic, Prague, Czech Republic\\
$^6$Department of Earth and Planetary Science, Graduate School of Science, The University of Tokyo, Tokyo, Japan\\
$^7$Dipartimento di Fisica, Universita della Calabria, I-87036 Cosenza, Italy\\
$^8$School of Physics and Astronomy, Queen Mary University of London, London E1 4NS, UK}
\end{center}

\begin{abstract}
We study the interaction of upstream ultra-low frequency (ULF) waves with collisionless shocks by analyzing the outputs of eleven 2D local hybrid simulation runs. Our simulated shocks have Alfv\'enic Mach numbers between 4.29-7.42 and their $\theta_{BN}$ angles are 15$^\circ$, 30$^\circ$, 45$^\circ$ and 50$^\circ$. The ULF wave foreshocks develop upstream of all of them. The wavelength and the amplitude of the upstream waves exhibit a complex dependence on the shock's M$_A$ and $\theta_{BN}$. The wavelength positively correlates with both parameters, with the dependence on $\theta_{BN}$ being much stronger. The amplitude of the ULF waves is proportional to the product of the reflected beam velocity and density, which also depend on M$_A$ and $\theta_{BN}$. The interaction of the ULF waves with the shock causes large-scale (several tens of upstream ion inertial lengths) shock rippling. The properties of the shock ripples are related to the ULF wave properties, namely thier wavelength and amplitude. In turn, the ripples have a large impact on the ULF wave transmission across the shock because they change local shock properties ($\theta_{BN}$, strength), so that different sections of the same ULF wave front encounter shock with different characteristics. Downstream fluctuations do not resemble the upstream waves in terms the wavefront extension, orientation or their wavelength. However some features are conserved in the Fourier spectra of downstream compressive waves that present a bump or flattening at wavelengths approximately corresponding to those of the upstream ULF waves. In the transverse downstream spectra these features are weaker.
\end{abstract}

\noindent {\bf Keypoints}
\begin{itemize}
\item Upstream ULF waves are not simply transmitted across collisionless shocks.
\item The appearance of the foreshock ULF waves changes as they cross collisionless shocks.
\item Some spectral features of compressive upstream ULF waves are conserved in the spectra of downstream fluctuations.
\item There is some correlation between the properties of the shock ripples and the foreshock ULF waves.
\end{itemize}

\section{Introduction}

The Earth's magnetosheath \citep[MSH, ][]{lucek:2005} is a region sandwiched between the bow-shock (BS) and the magnetopause of Earth. Its existence was first predicted by \cite{kellogg:1962}.
MSH is highly turbulent, perturbed by different low-frequency \citep[$\lesssim$ proton gyrofrequency, $\sim$0.01-0.1~Hz; ][]{song:1997, schwartz:1996}  fluctuations. 
Already at early times various authors \citep{mckenzie:1969, fairfield:1970, fairfield:1976, greenstadt:1970, barnes:1970} suggested that the magnetic field fluctuations in the MSH may be generated in various ways: in the magnetosheath itself, at the BS or in the foreshock region \citep{eastwood:2005}. 

There is a general consensus \citep[e.g. ][and the references therein]{schwartz:1996} that in the region of the MSH that lies downstream of the quasi-perpendicular BS ($\theta_{BN} >$ 45$^\circ$, where $\theta_{BN}$ is the angle between the upstream magnetic field and the local shock normal), also called quasi-perpendicular magnetosheath, the dominant wave modes are the Alfv\'en Ion-Cyclotron (AIC) and mirror mode (MM) waves which form either behind the BS or deeper in the MSH. The AIC instability is prevalent in the regions where the plasma $\beta$ (ratio between plasma thermal and magnetic pressures) is low ($<$1) while MM instability occurs in plasmas with $\beta\geq$1 \citep[e.g. ][]{gary:1976, gary:1993a, gary:1993b, tsurutani:1982, song:1992b, sckopke:1990, schwartz:1996, samsonov:2007}. 
In the quasi-parallel magnetosheath ($\theta_{BN} <$ 45$^\circ$), the wave activity is much higher. It is thought that fluctuations there are either transmitted foreshock waves or that they can be formed at the BS \citep[i.e. ][]{fairfield:1970, luhmann:1986, czaykowska:2001}. 
In the following sections we briefly summarize the state of the art regarding wave activity in the MSH.

\subsection{Waves in quasi-parallel magnetosheath}
The indirect evidence for the association of the MSH fluctuations with the bow-shock geometry came from observing that these fluctuations are not uniformly distributed in the MSH.
Dawn-dusk asymmetry in the long term average level of fluctuations, consistent with the idea that strong waves are associated with the quasi-parallel shock, was reported already by \cite{fairfield:1970}. The reason behind the dawn-dusk asymmetry lies in the average orientation of the interplanetary magnetic field (IMF) in the ecliptic which follows a nominal Parker spiral. It is more probable for the BS on the dawn side to exhibit quasi-parallel geometry, while on the dusk side the quasi-perpendicular configuration is more common.
\cite{luhmann:1986} were the first to search for a relationship between the spatial distribution of the MSH fluctuations and the IMF orientation and acknowledge the quasi-parallel BS to be an important contributor to the fluctuations in the dayside MSH. 

It is believed that upstream ultra-low frequency (ULF, periods between 10-100~s) waves from the foreshock may strongly influence the formation of downstream fluctuations. They may be partially transmitted into the downstream region and mode converted into downstream Alfv\'enic turbulence. Another source of downstream waves at quasi-parallel shocks is an interface instability that arises due to the interaction of incident ions and partially thermalized plasma at the shocks. These waves are only present near the shock transition region. \citep{kraussvarban:1991, kraussvarban:1995b, scholer:1997}.

\cite{czaykowska:2001} performed a study of fluctuations during 132 bow-shock crossings. They found that in the vicinity of quasi-parallel shocks the magnetic power was strongly enhanced, which the authors attributed either to wave generation at the shock or an amplification of convecting upstream waves at the shock interface.
\cite{du:2008} analyzed magnetosheath fluctuations in the 4-240~s range. They observed large-amplitude compressional and transverse fluctuations downstream of quasi-parallel shocks.
\cite{dimmock:2014} studied magnetic field fluctuations in the dayside MSH in the frequency range from 0.1-2~Hz. The authors showed a tendency of such fluctuations to exhibit higher amplitudes in the dawn flank magnetosheath and close to the magnetopause during southward IMF. 
A dawn-dusk asymmetry of Pc3 velocity fluctuations in the dayside MSH was shown to exist by \cite{dimmock:2016}. In general, larger amplitudes were observed downstream of the quasi-parallel BS and during the times of fast solar wind (SW). 

\subsection{Waves in the quasi-perpendicular magnetosheath}

It was suggested by, for example, \cite{omidi:1994} and \cite{mckean:1995}, that AIC and MM waves grow at or near quasi-perpendicular shocks and are then convected away by the sheath plasma. 
The fact that AIC and MM fluctuations are often observed downstream of quasi-perpendicular shocks is not surprising. These shocks heat the plasma preferentially in the perpendicular direction with respect to the local magnetic field \citep{winske:1988}, thereby enhancing the temperature anisotropy which is needed for the growth of these two types of waves. 

\cite{mckean:1995} found that as the short wavelength waves propagate downstream in the MSH, they are heavily damped so that in these regions only wave modes with longer wavelengths survive. The authors suggested that this is probably due to wave-particle scattering which results in gyrotropic and approximately bi-Maxwellian ion distributions.

It was shown by several authors that the temperature anisotropy T$_\perp$/T$_\parallel$ is inversely correlated with the parallel proton beta, $\beta_{\parallel, p}$ \citep{anderson:1994, fuselier:1994, phan:1994, gary:1993b, gary:1995}. 
\cite{fuselier:1994} argued that this inverse correlation is a consequence of pitch angle scattering of ions by electromagnetic ion cyclotron waves, which regulate the anisotropy and restore a marginally stable plasma.
\cite{chaston:2013} showed that broadband kinetic Alfv\'en waves heat magnetosheath ions. The authors observed that the energy density of ions correlates well with the energy density of these waves. The heating occurs predominantly in the direction perpendicular to the local magnetic field and is limited by the threshold condition for anisotropy instability.

\cite{hubert:1998} used data from ISEE 1 and 2 spacecraft to study the nature of low-frequency     waves during a crossing of the Earth's MSH downstream of a quasi-perpendicular ($\theta_{BN}$=51$^\circ$) bow-shock.  The authors observed a region of purely AIC waves in a 0.3~R$_E$ thick layer adjacent to the shock, followed by a region 2~R$_E$ thick where AIC and MM waves coexisted and finally a pure MM region. The authors thus argued that the dominant wave mode is controlled by the depth in the MSH.
\cite{czaykowska:2001} found that downstream of quasi-perpendicular shocks the magnetic wave activity is significantly enhanced compared to upstream. During times of low $\beta$ the authors found left-hand polarized AIC waves in the MSH. Strong correlation between the temperature anisotropy and the intensity of these waves was also found. Downstream of some highly oblique shocks, MM waves were also observed.

\cite{dimmock:2015b} showed that temperature asymmetry $T_\perp/T_\parallel>$1 is favored on the dusk (quasi-perpendicular) side and this is reflected in a similar asymmetry of mirror mode activity. T$_\perp$/T$_\parallel$ decreases with increasing SW Alfv\'enic Mach number (M$_A$), whereas mirror mode occurrence exhibits the opposite trend. The dawn-dusk asymmetry diminishes with increasing M$_A$. Also, during the transition from low to moderate M$_A$ the authors observed a shift in the data from MM dips to MM peaks.

\cite{soucek:2015} presented a statistical study of the spatial distribution of MM and AIC waves in the MSH as a function of relevant plasma parameters, such as ion temperature anisotropy and ion $\beta$. The authors showed a strong dependence of the two plasma parameters and the occurence of the waves on the shock's $\theta_{BN}$ and its M$_A$. It was found that the AIC waves occur almost exclusively in the plasma that is stable to MM instability. MM were found to occur in the quasi-perpendicular MSH in correlation with high M$_A$, while lower M$_A$ favour the AIC waves. Both are rarely observed in the quasi-parallel MSH. AIC are observed behind the shock in the MSH flanks and less so near the subsolar MSH.

\subsection{Wave transmission across the bow-shock}
Although plenty of works have been published on magnetosheath fluctuations, questions such as what happens to the foreshock waves as they cross the BS and how the properties of downstream fluctuations relate to those of the foreshock waves, are still not completely answered. 

It has been shown that the upstream ULF waves are a mixture of Alfv\'en and magnetosonic waves \citep[e.g. ][]{sentman:1981, hoppe:1983, eastwood:2003}. Some of the early works \citep[][]{mckenzie:1969, mckenzie:1970a, mckenzie:1970b, asseo:1970} concluded that the magnetosonic and Alfv\'en waves are strongly amplified on passage through the shock. On the other hand, it was found by \cite{mckenzie:1974}, that the waves also impact the shock: Alfv\'en waves diminish the shock compression ratio while sound waves either enhance or diminish the compression ratio depending on whether the incident wave vector is parallel or antiparallel to the upstream flow direction.  \cite{asseo:1970} showed that when an incident wave strikes a shock front, it is refracted and gives rise to five other hydromagnetic waves. \cite{hassam:1978} concluded that small amplitude Alfv\'en waves may perturb the shock surface and give rise to transmitted waves consisting of a fast magnetosonic wave, forward and backward slow magnetosonic waves and an entropy wave.
\cite{whang:1987} showed that the number of downstream waves excited due to transmission of the upstream waves depends on the angle of incidence: when this angle is lower than some critical value, six diverging downstream waves are excited. When the angle of incidence is larger than a crititial value, the number of downstream excited waves may be smaller than six. \cite{lu:2009} used 2D hybrid simulations in order to study the interaction of Alfv\'en waves with a quasi-perpendicular shock. The authors found that the Alfv\'en waves are transmitted through the shock and that their amplitude is enhanced 10-30 times. The authors also found that the shock ripples form due to the upstream Alfv\'en waves.

There have been some works showing indirect evidence in favor of the upstream ULF wave transmission into the magnetosheath and even into the magnetosphere. It has been known since roughly 50~years now that a subset of the magnetospheric ULF waves, dayside Pc3-4 pulsations, is enhanced during radial IMF configurations \citep{troitskaya:1971}. This has been explained in terms of the foreshock ULF waves being the source of the magnetospheric waves in the same frequency range. Several studies have observed simultaneously waves of similar frequencies in the foreshock region and in the magnetosphere \citep[e.g. ][]{greenstadt:1983, russell:1983c, engebretson:1991, lin:1991, clausen:2009, villante:2011, francia:2012, takahashi:2016, takahashi:2021}. Some local \citep{kraussvarban:1991} and global \citep{shi:2013, shi:2017} hybrid simulations also favour upstream ULF wave transmission into the magnetosheath and the magnetosphere. On the other hand \cite{narita:2005}, \cite{narita:2006} and \cite{narita:2006b} do not favour the wave transmission across the bow-shock. Their arguments are that from the viewpoint of the dispersion relation and other wave properties, such as propagation angle and  polarization, the foreshock waves are not transmitted into the magnetosheath.

Spacecraft observations alone cannot provide a complete answer to the question of what happens to the waves as they cross the shock, since such data are spatially limited. However local and global hybrid simulations of upstream wave--shock interaction may provide us with some clues on how this interaction occurs. Local simulations (with self-consistent foreshock) are complementary to global simulations with complex foreshock which links regions of shock with different $\theta_{BN}$.
In this work we try to answer this question by analysing the outputs of local 2.5D hybrid (kinetic ions, fluid electrons) simulations of collisionless shocks and the corresponding upstream and downstream regions. In total we show results from eleven local runs with Alfv\'enic Mach numbers M$_A$ ranging between 4.29 and 7.42 and the $\theta_{BN}$ between 15$^\circ$ and 50$^\circ$. Thus we simulate quasi-parallel and marginally quasi-perpendicular shocks. These shock properties are comparable to those of the Earth's bow-shock. In the case of the latter the typical M$_A$ at its nose ranges between 6 and 7 \citep{winterhalter:1988} while it is lower towards the flanks.

In the remainder of this paper we first describe the simulation setup in Section~\ref{sec:setup}. Next, we discuss the simulation results in Section~\ref{sec:results}. In this section we begin by analyzing  the degree of the shock rippling at different shocks. We emphasize this process before studying the downstream ULF wave properties since, as we show later on, shock rippling is crucial for understanding how the ULF waves are affected by the shocks. This process results from the fact that the shock reformation is not in phase all over the shock surface. Thus, the shock properties ($\theta_{BN}$, compression ratio, etc.) change with the location on the shock surface, so different sections of a single upstream ULF wave encounter the shock with different properties, which then affects the wave transition. In the Section~\ref{sec:casestudy} we analyze an example of a ULF wave interacting with a shock. Finally in the Section~\ref{sec:discussion} we discuss the results of this work and present the conclusions.

\section{Simulation setup}
\label{sec:setup}
Local hybrid simulations were performed with the 2.5D HYPSI numerical code \citep{burgess:2015, gingell:2017, trotta:2019}. In these simulations, the electrons are treated as a massless, charge-neutralizing fluid, while ions, in our case pure protons, are treated kinetically. We perform a series of simulations with a grid of $N_x \times N_y$ = 1000 $\times$ 800 cells whose dimensions are of 0.5~d$_i$ (d$_i$ is the upstream ion inertial length) in both directions, respectively. The SW is injected from the left along the $x$-axis with inflow velocities of 3.3~V$_A$, 4.5~V$_A$ and 5.5~V$_A$ (V$_A$ is the initial Alfv\'en speed). The upstream magnetic field lies in the XY plane. Its initial magnitude is one and its orientation is such that four different shock geometries are produced with $\theta_{BN}$ = 15$^\circ$, 30$^\circ$, 45$^\circ$ and 50$^\circ$. The upstream ion and electron $\beta$ (ratio between thermal and magnetic pressures) are 0.5. Initially there were 100 particles per cell. The simulation is periodic in $y$ direction, while the right boundary acts as a reflective wall. Time is measured in units of the inverse of proton gyrofrequency ($\Omega^{-1}$). The simulation timestep is 0.005~$\Omega^{-1}$ and the outputs are produced every 2.5~$\Omega^{-1}$. In total we performed eleven simulations with initial conditions and the resulting shock Mach numbers that are summarized in Table~\ref{tab:hypsi}. As the incident protons are reflected at the right boundary, a shock is created which then starts to propagate towards the left. Throughout the paper the physical quantities, such as velocity and dynamic pressure, are calculated in the shock-rest frame.
The running times of the simulations range between and 250-325~$\Omega^{-1}$.

\section{Results}
\label{sec:results}
\subsection{Overview}
Figures~\ref{fig:bz} and \ref{fig:bf} show simulation outputs for all our runs at times when the bow-shock was located at $x\sim200~d_i$ in the simulation domain. The upstream regions lie to the left of the shocks. Figure~\ref{fig:bz} exhibits maps of B$_z$, while Figure~\ref{fig:bf} shows $|$B$|$. The units of magnetic field are normalized to the initial $|$B$|$ value.

The different panels show shocks with different properties. The $\theta_{BN}$ increases to the right, while the inflow velocity and thereby the M$_A$ of the shocks increases from top to bottom. We also observe that M$_A$ increases slightly with the increasing $\theta_{BN}$. This is due to the fact that for the same upstream SW properties (density, velocity, temperature), the quasi-perpendicular shocks tend to propagate faster than the quasi-parallel ones.

The colorscales in both figures are centered on the initial upstream values. 
Black lines mark the cuts along which the wave properties are studied in Section~\ref{sec:thewaves} and were selected so as to include as many wavefronts as possible.
We first look at the B$_z$ plots shown in Figure~\ref{fig:bz}. We can see that the wavelength and the orientation of the wave fronts  of the upstream fluctuations varies from panel to panel. B$_z$ fluctuations are by definition transverse, which means that they propagate at small angles with respect to the magnetic field direction and their wavefronts are approximately perpendicular to {\bf B}. The wavefronts in the first column (panels a, d, h), where the magnetic field makes an angle of 15$^\circ$ with respect to the $x$-axis, are oriented almost perpendicular to $x$. On the panels in the last column, where nominal $\theta_{BN}$=50$^\circ$ the upstream wavefronts are strongly inclined with respect to the $y$-axis.

The properties (wavelength and the amplitude) of the upstream waves change with M$_A$ (from top to bottom) and $\theta_{BN}$ (from left to right), which is addressed in more detail in the Discussion section.

In the case of the $|$B$|$ fluctuations (Figure~\ref{fig:bf}), the upstream wave fronts are much more aligned with the magnetic field. The analysis of sequential outputs reveals that these waves propagate obliquely to B. It was shown by \cite{omidi:2007b} and \cite{blancocano:2011} that these fluctuations are compressive fast magnetosonic waves. We can see in Figure~\ref{fig:bf} that in the case of waves on panels b), c), g) and k), the wavefronts differ from other runs in that they do not seem to be always aligned with the magnetic field but many of them appear like more localized magnetic field enhancements.

Downstream of the shocks the fluctuations exhibit a very different appearance. In all cases a filamentary structure can be observed. The orientation of these filaments is different from that of the upstream waves and changes with the increasing nominal $\theta_{BN}$: as this angle increases, the orientation of the filaments forms an increasingly larger angle with respect to the direction of the nominal shock normal.%
\subsection{Statistical analysis of the shock ripples}
Figures~\ref{fig:bz} and \ref{fig:bf} show that the wavefronts of the upstream waves extend several tens of d$_i$ to $\gtrsim$100s d$_i$. On the other hand, the simulated shocks exhibit surfaces which are rippled and variable in time. Both facts influence importantly the wave transmission across the shocks. We show a close-up of one of the shocks in Figure~\ref{fig:bf089}. Here colors represent the magnetic field magnitude, the vertical magenta line marks the average shock location (here the coordinate system is shifted along the $x$-axis so that this location is at $x$=0. Note that the Figure only shows a portion of the simulation domain, which extends further in the $y$ direction. There are other ripples, which result in the average shock position being centred on zero in this plot.) and the thick black line represents the shock surface. The latter was calculated the following way:
\begin{enumerate}
    \item First, we average the magnetic field profile in $y$ direction at the time t=225.5~$\Omega^{-1}$. Then we determine the $x$-position of the shock to be where the averaged B-magnitude first reaches the value of 2.5. This value was selected since it coincides with the shock transition in the averaged shock profiles.
    \item Afterwards we determine the position of the shock surface in the original (not averaged) magnetic field data. For each $y$-coordinate we search for the first $x$-coordinate at which the B-magnitude reached the value of 2.35. The search was done in the vicinity of the shock position calculated from the averaged shock profile. The value of 2.35 was selected by eye after many attempts to approximate each individual shock surface with a single curve. The reason why the search for this value was made near the position of the shock determined from the average B-field profile, was to avoid the upstream compressive structures, such as shocklets, to be identified as parts of the shock.
    \item Next, we smooth the calculated $y$-coordinates using the Savitzky-Golay filter with width of 21 points and of the third order. This calculated surface can be seen in Figure~\ref{fig:bf089} as a thick black line. We can see that the curve closely matches the magnetic field jump.
    \item Finally, we average all $x$-values of this black line to obtain an averaged shock location. This location is different from that calculated from the averaged magnetic field profile and this is the position marked by the purple vertical line in Figure~\ref{fig:bf089}.
\end{enumerate}

In order to establish how rippled the shocks are we show three diagnostics that describe three properties of the ripples: the amplitudes, the length scales and how steep their sides are.

In Figures~\ref{fig:figure4}a) and b) we show two histograms that exhibit the distributions of distances ($\Delta x$) for every point on the shock surfaces from the corresponding shock's average location. The panel on the left exhibits the histogram for the shock from the run {\it a} (from now on we will refer to different runs with the letters that correspond to the panels in the Figures~\ref{fig:bz} and \ref{fig:bf}), while the one on the right is for the shock from the run {\it i} (histograms for all the runs can be seen in Figure~\ref{fig:appendix1} in the Appendix). The most extreme values in each histogram, whether negative or positive, provide information about the maximum amplitudes of the ripples. The width of the histograms (represented by $\sigma$) also provides information on the ripples: small $\sigma$ means that the majority of the ripples have smaller amplitudes, while large $\sigma$ point to ripples with larger amplitude. We can see on panel \ref{fig:figure4}a), that the bulk of the $\Delta$x values is below 10~d$_i$. The distribution is strongly peaked at around 0 and the standard deviation is $\sigma(x)$=3.94~d$_i$, the smallest among all runs. The value of $\sigma$ for the run {\it i} (\ref{fig:figure4}b) is 8.76~d$_i$, the largest of all. Figure~\ref{fig:figure5}a) exhibits $\sigma$ values for all the runs. The horizontal dashed line marks the value of $\sigma$=6.0. We divide the simulated shocks into those with $\sigma\leq$6.0 (from runs {\it a, b, f, h, j} and {\it k}) and those with $\sigma>$6.1 (from runs {\it c, d, e, g, i}). This division is arbitrary but it will help us with further analysis.

Figures~\ref{fig:figure4}c) and d) show power spectra of the curves describing the surfaces of the shocks from runs {\it a} and {\it i}. An example of such a curve is shown in Figure~\ref{fig:bf089}. The idea is to check whether the ripples exhibit specific wavelengths. 
Only two spectra are exhibited in this Figure, however spectra for all shocks are presented in Figure~\ref{fig:appendix2} in the Appendix. All of them are continuous and their power at wavelengths above 40~d$_i$ (vertical blue lines) is enhanced compared to the trend at smaller wavelengths (red dashed lines). On the panels \ref{fig:figure4}c) and d) we state a quantity called ``integrated power'' which was obtained by first integrating the spectra and then subtracting the power that can be attributed to small-scale (below 40~d$_i$) background (red dashed lines on panels \ref{fig:figure4}c and d). We can see that this value is 16.5 for the run {\it a} (smallest among all) and 50.4 for run {\it i} (largest of all).
Again, based on Figure~\ref{fig:figure5}b), which shows integrated power values for all shocks, we can divide the shocks into two groups - runs {\it a, b, c, f, g, h, j} and {\it k} exhibit integrated power$<$26 (horizontal dashed line), while the runs {\it d, e} and {\it i} exhibit the integrated power$>$26.

Figures~\ref{fig:figure4}e) and f) show another statistic from the simulations. For each point on the shock surface we calculate the local normal at that position. Then we calculate the angle between every possible pair of normals, $\theta_{NN'}$, and plot this as a function of distance between the points (plots for all the runs are exhibited in the Figure~\ref{fig:appendix3} in the Appendix). The maximum angles observed for each case are a proxy of how strongly the shocks are rippled.

We can make a thought experiment in order to have a clearer picture about what the plots in Figures~\ref{fig:figure4}e) and f) tell us. If the shock surface could be described with a single sinusoidal curve and if we took one point on a crest of one of the ripples, and plotted $\theta_{NN'}$ values as a function of a distance from that point, we would clearly observe $\theta_{NN'}$ peaks and dips at regular intervals corresponding to the half of the wavelength of that sinusoidal function. $\theta_{NN'}$ would exhibit values of 0 at the locations of the crests (minima or maxima since the shock normals by definition always point towards upstream). If we used more than one sinusoidal function to describe the shock surface, the picture would become more complicated - we would observe many irregular peaks at intervals that would exhibit some quasi periodicity. We would also get a range of $\theta_{NN'}$ values at each distance. If the typical wavelength of the ripples was small, we would get many thin peaks, if it was large, fewer and thicker peaks would be produced. $\theta_{NN'}$ values close to 180$^\circ$ would only be obtained between points lying on the sides of steep ripples, pointing in opposite $y$ directions.

We can see from Figures~\ref{fig:bf}a), i) and \ref{fig:figure4}e), f) that there seems to be a correlation between the upstream compressive waves and the peaks : in the case of the run {\it i} the wavefronts are many and exhibit small widths, and this is likely why the Figure~\ref{fig:figure4}e) exhibits many peaks which tend to be narrower. On the contrary, wider and less numerous upstream waves, such as those in Figure~\ref{fig:bf}i), are associated with wider and fewer peaks (Figure~\ref{fig:figure4}f). 

We show three numbers on the panels \ref{fig:figure4}e) and f): the maximum and median $\theta_{NN'}$ and $S$, which is the fraction of the area on the panels delimited by the black points.
Once again, based on the Figures~\ref{fig:figure5}c) and d), we can divide the runs in two groups: the runs {\it a, b, f, h, j} and {\it k} exhibit median $\theta_{NN'}$ values $\leq$79$^\circ$ and $S\leq$0.42 (horizontal dashed lines on both panels), while the the runs {\it c, d, e, g, i} exhibit higher values of $\theta_{NN'}$ and $S$.

We can see that the shocks from the runs {\it a, b, f, h, j} and {\it k} all form a group that consistently exhibits smaller $\sigma$, integrated power, $\theta_{NN'}$ and $S$ values. Just for illustrative purposes, we call these shocks weakly rippled. The shocks from the runs {\it d, e, i} always exhibit larger values, so we refer to them as strongly rippled. The shocks from the runs {\it c} and {\it g} exhibit higher values of $\sigma$, $\theta_{NN'}$ and $S$, but smaller values of the integrated power. Hence, we denominate them as intermediately rippled.

\subsection{Analysis of upstream and downstream wave properties}
\label{sec:thewaves}
\subsubsection{Fourier spectra}
Figures~\ref{fig:figure4}g) and h), i) and j) show Fourier spectra for upstream (black curve) and downstream (red) waves. Spectra from all runs can be seen in Figures~\ref{fig:appendix5} and \ref{fig:appendix6}. These are instantaneous spectra obtained at simulation times indicated in Figures~\ref{fig:bz} and \ref{fig:bf} along the cuts marked with black lines. These cuts were chosen so that they are aproximately perpendicular to the wave fronts and they include the maximum number of the waves. This way the units on the $x$-axis in Fourier spectra represent true  wavelengths, not some projections along the cuts and the power of the foreshock in the Fourier spectra waves is maximized. The spectra were obtained by first windowing the original data with the Tukey (tapered cosine) function \citep{harris:1978} and then applying a 13-point running average filter on the FFT power spectra.

We can see that the downstream spectra exhibit higher power than upstream spectra. The upstream spectra of transverse (B$_z$) and compressive ($|$B$|$) fluctuations are rather flat at wavelengths $\lesssim$10-60~d$_i$. The exact value depends on the run. We shall refer to these intervals as ``ULF range'' (shaded in blue) since the simulated upstream waves are analogous to the foreshock ULF waves. Parts of the upstream spectra at wavelengths above the ULF range tend to increase smoothly without strong features. 

Normally we expect the foreshock ULF waves to produce a bump in the Fourier power spectra \citep[e.g., ][]{greenstadt:1995,hoppe:1982,hoppe:1983,le:1992,le:1994}. However in  these studies the data contain many wavefronts. In our case, as can be seen in Figures~\ref{fig:bz} and \ref{fig:bf}, the upstream cuts include $\lesssim$10 wavefronts, in some cases as few as three (see for example panels \ref{fig:bz}h and \ref{fig:bf}k). Due to the fact that the number of wavefronts is low, the small peaks in Fourier spectra do not merge into a single bump. 

We can observe certain features in the upstream spectra that are also present downstream. In the case of the weakly rippled shock from the run {\it a} (\ref{fig:figure4}g and i) the flattening can be clearly observed in the $|$B$|$ and B$_z$ spectra. In the case of the strongly rippled shock from the run {\it i}, this flattening is much less obvious in the B$_z$ spectrum than in the $|$B$|$ spectrum. In general (see Figures~\ref{fig:appendix5} and \ref{fig:appendix6}), flattening in the ULF range can be observed in the case of the spectra shown on panels a), b), f), h), j), and  k), which are classified as weakly rippled. In the case of the panel g), the downstream spectrum does not exhibit flattening but several peaks. In the case of the runs d) and i) the downstream spectra are rather featureless. It thus seems that flattening in the ULF range is better observed downstream of weakly rippled shocks.

In general the downstream spectra are more turbulence-like than the upstream spectra meaning that the non-turbulent features (bumps and flattening) are less prominent. The latter holds even more for the transverse, B$_z$ spectra in Figure~\ref{fig:figure4}i) and j). 

\subsubsection{Wavelengths and amplitudes of upstream waves}
\label{sec:theory}
In previous sections we imply that the upstream wave properties determine how they are going to be affected during shock crossing. Their wavelengths and amplitudes determine the properties of the shock ripples, which in turn affect the transmitted wave properties. Here we take a look on how the wavelength and the amplitude of the waves depend on M$_A$ and $\theta_{BN}$.

We have seen from the Figures~\ref{fig:bz} and \ref{fig:bf} that the wavelength of the upstream waves varies with M$_A$ and $\theta_{BN}$, however the correlation with these two parameters is not obvious at the first sight. 
\cite[][]{watanabe:1984} have shown that the frequency of the upstream ULF waves in the spacecraft frame, $\omega_s$, depends on the upstream solar wind speed V$_{SW}$, the reflected ion beam speed, V$_b$, the wave phase speed V$_{ph}$ and angles $\theta_{kV}$ and $\theta_{kB}$, between the wave vector $\vec{k}$ and the upstream SW velocity and B-field directions, respectively. These authors derived the expression for the ratio between $\omega_s$ and the proton gyrofrequency, $\Omega_p$:

\begin{equation}
    \frac{\omega_s}{\Omega_p} = \Bigg|\frac{V_{ph} + V_{SW}\cos\theta_{kV}}{V_{ph} - V_b\cos\theta_{kB}}\Bigg|.
    \label{eq:watanabe84}
\end{equation}

In order to obtain the V$_b$ one needs to know the origin of the upstream ULF waves. It was first proposed by \cite{fairfield:1969} that these waves are formed due to the electromagnetic ion beam-cyclotron instability excited by the reflected, upstream propagating field-aligned ion beams. This instability generates low-frequency, right-hand mode waves. 
In order for them to be excited, a cyclotron resonance condition must be fulfilled:
\begin{equation}
    \omega - \vec{k}\cdot\vec{V'_b} = -\Omega_p,
\end{equation}
where $\omega$ and V$'_b$ are the wave frequency and the beam velocity in the rest frame of the upstream plasma. 
The beam velocity in the spacecraft frame can be calculated by following \cite{schwartz:1983} calculations for magnetic moment-conserving reflection at the shock:
\begin{equation}
    V_b = V_{SW}\bigg(2\frac{\cos\theta_{VN}}{\cos\theta_{BN}} - 1\bigg)^{1/2}.
    \label{eq:beamvelocity}
\end{equation}

We can now start putting the pieces together. Since the upstream ULF waves in our simulations are mostly transverse, we may approximate their phase velocity with the upstream Alfv\'en speed, $V_{ph}\sim V_A$. Their propagation direction and thus their wave vectors $\vec{k}$ are approximately parallel to the upstream magnetic field, thus yielding $\theta_{kB}\sim0$ and $\theta_{kV}\sim\theta_{BV}$. In our simulations, the ``spacecraft rest frame'' corresponds to the shock normal incidence frame in which the upstream SW velocity and the shock normal anti-parallel. Thus the acute angle $\theta_{VN}$ (betwen the SW velocity and shock normal) is 0 and $\theta_{BV} = \theta_{BN}$. This, plus the expression~(\ref{eq:beamvelocity}) may be introduced into the expression~(\ref{eq:watanabe84}) to obtain
\begin{equation}
    \frac{\omega_s}{\Omega_p} = \Bigg|\frac{V_A + V_{SW}\cos\theta_{BN}}{V_A - V_{SW}\big(\frac{2}{\cos\theta_{BN}} - 1\big)^{1/2}}\Bigg|.
\end{equation}
Dividing the upper and lower parts of this equation by V$_A$ we obtain:
\begin{equation}
    \frac{\omega_s}{\Omega_p} = \Bigg|\frac{1 + M_A\cos\theta_{BN}}{1 - M_A\big(\frac{2}{\cos\theta_{BN}} - 1\big)^{1/2}}\Bigg|.
\end{equation}
By taking into account that $\lambda = 2\pi V_{ph}/\omega_s$, we arrive to the expression for the wavelength of the upstream waves:
\begin{equation}
    \lambda = \frac{2\pi V_A}{\omega_s} = \frac{2\pi V_A}{\Omega_p} \Bigg|\frac{1 - M_A\big(\frac{2}{\cos\theta_{BN}} - 1\big)^{1/2}}{1 + M_A\cos\theta_{BN}}\Bigg|.
\end{equation}

This wavelength does not depend on the selected frame of reference and thus corresponds to the wavelength measured in the rest frame of the simulation domain. The numeric factor $\frac{2\pi V_A}{\Omega_p}$ is the same for all our simulations, so we can define the normalized wavelength as $\lambda$ divided by this factor. This quantity is shown in Figure~\ref{fig:figure13} for all our models. We can see that the quantity that most strongly correlates with the normalized $\lambda$ is the $\theta_{BN}$. The normalized $\lambda$ for models h)-k) (V$_{in}$=5.5~V$_A$) varies between $\sim$0.8 and $\sim$1.7, so by a factor of $\sim$2.1. Maintaining the $\theta_{BN}$=50$^\circ$ (models c), g) and k)) and varying the M$_A$ by varying the inflow velocity V$_{in}$, produces normalized $\lambda$ values between 1.43 and 1.70, so only by a factor of $\sim$1.2.
By visual inspection of the Figure~\ref{fig:bz}, we can observe that the wavelength of the upstream waves indeed increases strongly with $\theta_{BN}$ (from left to right), while its increase due to V$_{in}$ (from top to bottom) is somewhat smaller, although still significant.

Another relevant property that changes with varying upstream conditions is the amplitude of the upstream waves. The wave amplitude may influence the amplitude of the magnetic field amplitude of the shock ripples. It has been shown by \cite{barnes:1970} that the amplitude of upstream ULF waves is proportional to the product of the beam velocity and density:
\begin{equation}
    \big<\delta B^2\big>\propto V_bN_b.
\end{equation}
We know from equation~\ref{eq:beamvelocity} that for the acute angle $\cos\theta_{VN}$=0, the V$_b$ increases with increasing $\theta_{BN}$.
Similar conclusion was reached by \cite{burgess:1987}. These authors, by employing 1D hybrid simulations, showed that i) the field-aligned beam velocity increases and ii) its density decreases strongly with increasing $\theta_{BN}$ for 40$^\circ<\theta_{BN}<$60$^\circ$.
This interplay between V$_b$ and N$_b$ is probably the reason why the amplitudes of the upstream waves exhibit a complicated dependence on the shock's M$_A$, $\theta_{BN}$ in Figures~\ref{fig:bz} and \ref{fig:bf}. 

\section{Case study}
\label{sec:casestudy}
The spectra presented in the previous section suggest that as the upstream waves cross the shock their properties are at least partially modified. Here we look at a sequence of outputs produced by the run {\it e} with M$_A$ = 7.08 and $\theta_{BN}$ = 30$^\circ$ in order to analyse what happens to the waves as they encounter the shock. This run was chosen since it produces a strongly rippled shock.
Figures~\ref{fig:bf30} and \ref{fig:bz30} exhibit a sequence of excerpts of outputs produced at times indicated on each panel. The former Figure shows the magnetic field magnitude while the latter exhibits the B$_z$ component. 

We first look at the Figure~\ref{fig:bf30}. It is quite obvious that the shock front is rippled and its appearance changes between panels. On panel a) we observe a compressive structure in the upstream region, shaded with the light blue color, located approximately between y=(-20, 30)~d$_i$. This structure, the equivalent of steepened ULF waves or compressive structures such as shocklets, is already interacting with the shock at y$\sim$(-0, 20)~d$_i$. Short wavelength fluctuations, equivalent of the whistler waves at real shocks, are present close to this interaction region.

At this location the shock is locally concave (yellow arrows), i.e. this portion of the surface is located slightly to the right compared to the rest of it. By t=225~$\Omega^{-1}$ (panel b) the region of the most intense interaction between the compressive upstream structure and the shock expanded to y$\sim$(-19, 35)~d$_i$. The elevated magnetic field magnitude in this section, marked with the white trace between y$\sim$(-3, 30), means that the upstream structure had steepened close to the shock. This interaction region is also permeated by small wavelength fluctuations, equivalent to the whistler mode waves commonly observed upstream of collisionless shocks.
By t=230~$\Omega^{-1}$ (panel c) almost all of the upstream structure had passed into downstream. The shock is now much more locally concave than during previous times. 
At y$<$-20~d$_i$ the shock is convex and there is another upstream structure approaching it. We can see in panels d) and e) that the story now repeats on somewhat smaller scale as the shock surface between y$\sim$(-20, -40)~d$_i$ becomes concave.

We can see on panels a), b) and c) that as different sections of the arriving wave interact with the shock, the concave section of the shock surface propagates along the shock front. In this case it moves downward.

The transverse upstream fluctuations approaching the shock are shown in Figure~\ref{fig:bz30}. The colors represent the B$_z$ component. It can be seen that these fluctuations do not cause the shock rippling but are affected by it as different portions of a single wavefront reach the shock at different times and locations along the shock surface. Since the shock properties, such as local normal orientation and the shock strength, vary with location, the interaction of different portions of upstream waves is different and this leads to fragmentation of waves as they cross the shock.

The shock transition region (dark red trace at and immediately behind the shock transition in Figures~\ref{fig:bf30} and \ref{fig:bz30}) seems fairly complex. Regions of different magnetic field strength are mixed and tend to be oriented vertically, except behind the ripples, where they look much more disturbed. The wavefronts do not seem to conserve their shapes in these regions.

Farther downstream the magnetic field magnitude is lower than in the shock transition layer. Locations of magnetic field magnitude of different strength (blue, white, orange and red traces) are mixed together and exhibit filamentary structure with their orientations becoming increasingly vertical further downstream. They do not look similar to the shapes of the upstream waves. They exhibit much smaller $\lambda$, their sizes are smaller and their orientation is different. This is true for $|$B$|$ and B$_z$ fluctuations.

\section{Discussion and Conclusions}
\label{sec:discussion}
We performed eleven 2.5D local hybrid simulations of collisionless shocks in order to study the interaction of upstream ULF waves with the shocks. The shock's Alfv\'enic Mach numbers range between 4.29 and 7.42 while their nominal angles $\theta_{BN}$ = 15$^\circ$, 30$^\circ$, 45$^\circ$ and 50$^\circ$. We divide our study in two parts: we characterize the degree of rippling of the simulated shocks in the first part and the properties of the magnetic fluctuations in the second.

We acknowledge that our shocks are much simpler than real ones, such as the bow-shock of Earth. For one, in the case of the bow-shocks, the upstream ULF waves may be a mix of waves that are produced upstream of the sections of the bow-shock that have different nominal geometries and Mach numbers and produce waves with different properties. Also, due to the draping of the IMF lines around the magnetopause the regions deep in the magnetosheath may be magnetically connected to the regions of the bow-shock that are much farther away than is the case in our simulations. Another difference is the propagation direction of the downstream waves. For example, \cite{narita:2006} and \cite{narita:2006b} showed that magnetosheath low frequency waves propagate toward the flank magnetosheath for small zenith angles (close to the Sun-Earth line), while they propagate towards the magnetopause for large zenith angles. This ordering will not be observed in local simulations. Finally, in the case of planetary bow-shocks, the waves have much more time to grow in the corresponding foreshock regions. They also develop over larger distances from the bow-shock before being convected back to the shock.

Past works have shown that besides the ULF waves, there are also transient structures in the foreshock, such as foreshock cavitons \citep[e.g. ][]{blancocano:2009, blancocano:2011, kajdic:2011,kajdic:2013} and spontaneous hot flow anomalies \citep[SHFAs, ][]{zhang:2013, omidi:2013b}. The latter form when the foreshock cavitons approach the bow shock which results in further local ion heating and plasma and magnetic field depletion. \cite{omidi:2014} showed that SHFAs form for M$_A\geq$5. {\bf These structures contribute to the shock rippling. To our best knowledge, the relative importance of SHFAs versus ULF waves has not yet been quantified. \cite{kajdic:2013} showed that spacecraft on average observe $\sim$2 cavitons per each 24 hours spent in the foreshock. Hybrid simulations \citep[e.g. ][]{omidi:2007b} indicate that for each caviton detected by the spacecraft several more are formed that go undetected but that also contribute to shock rippling. On the other hand, the observations reported by \cite{kajdic:2013} show that for each foreshock caviton, tens or even hundreds of compressive ULF waveforms are detected. This may indicate that the ULF waves influence shock rippling more than the SHFAs. Here we do not specifically search for the cavitons and SHFAs, however eight of our eleven shocks have M$_A>$5. The visual inspection of the Figure~\ref{fig:bf} does not reveal any obvious candidates for SHFAs. The apparent absence of these structures may be due to the fact that in our simulated shocks the upstream region is quite limited in size, so that the compressive ULF waves have less time to grow than is the case for the real bow-shock. This results in lesser steepening of these waves. Since according to \cite{omidi:2007b} the foreshock cavitons form due to nonlinear interaction of compressive and transverse ULF waves, less wave steepening may result in a smaller number of SHFAs in our simulations.}

\begin{landscape}
\begin{figure}[h]
\centering
\includegraphics[height=1.0\textheight]{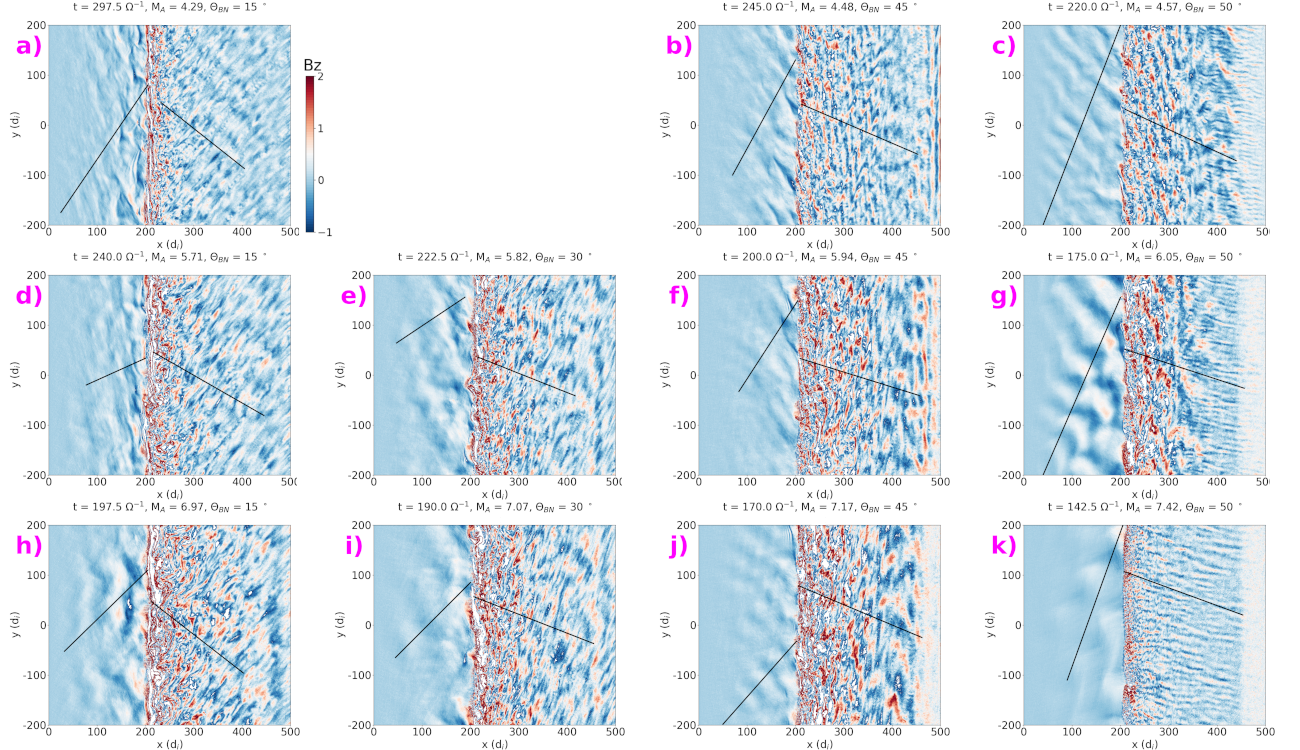}
 \caption{Simulation results of eleven local hybrid runs at times when the average shock location was x$\sim$200~d$_i$. Colors represent the B$_z$ component and the two black lines mark the locations along which the Fourier spectra in Figures~\ref{fig:figure4}g)-k) are calculated. $\theta_{BN}$ increases from left to right and M$_A$ from top to bottom.}
 \label{fig:bz}
 \end{figure}
\end{landscape}

\begin{landscape}
\begin{figure}[h]
\centering
\includegraphics[height=1.0\textheight]{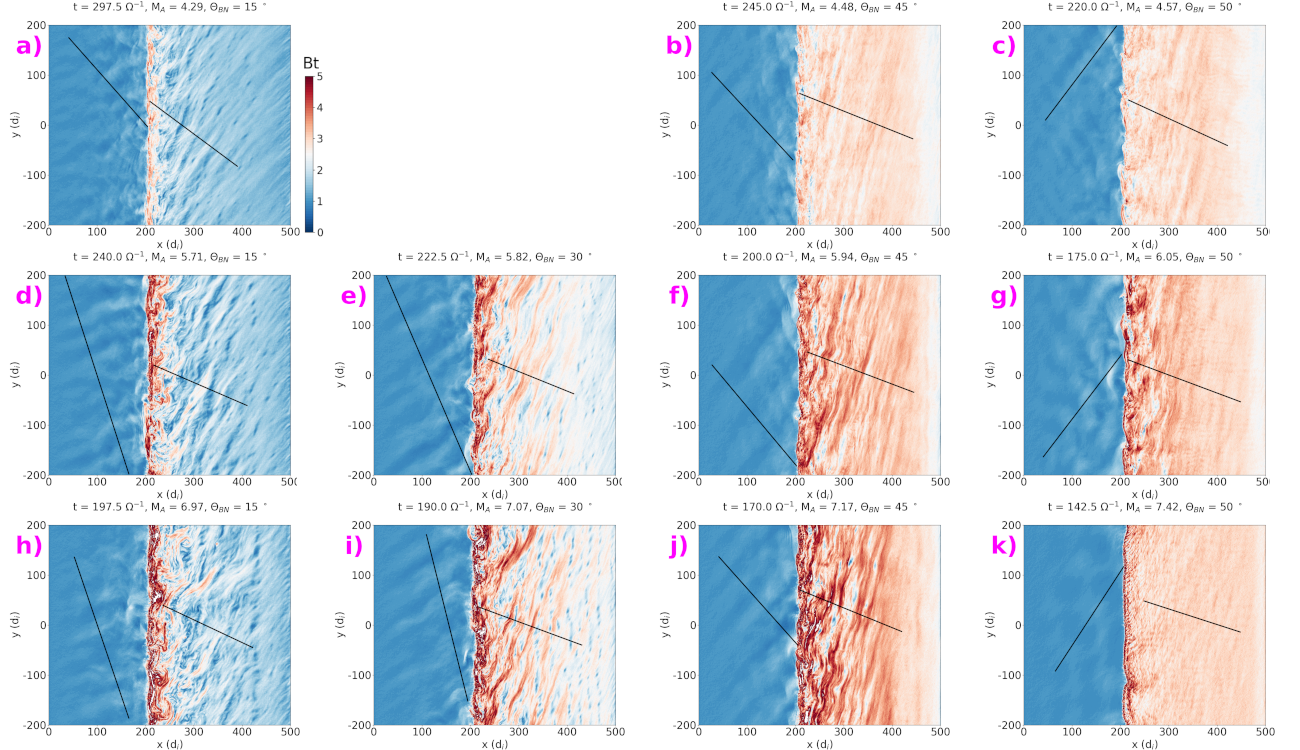}
 \caption{Simulation results of eleven local hybrid runs at times when the average shock location was x$\sim$200~d$_i$. Colors represent the B magnitude and the two black lines mark the locations along which the Fourier spectra in Figures~\ref{fig:figure4} and \ref{fig:appendix5} are calculated.}
 \label{fig:bf}
 \end{figure}
\end{landscape}

\begin{figure}[h]
\centering
\includegraphics[height=0.5\textheight]{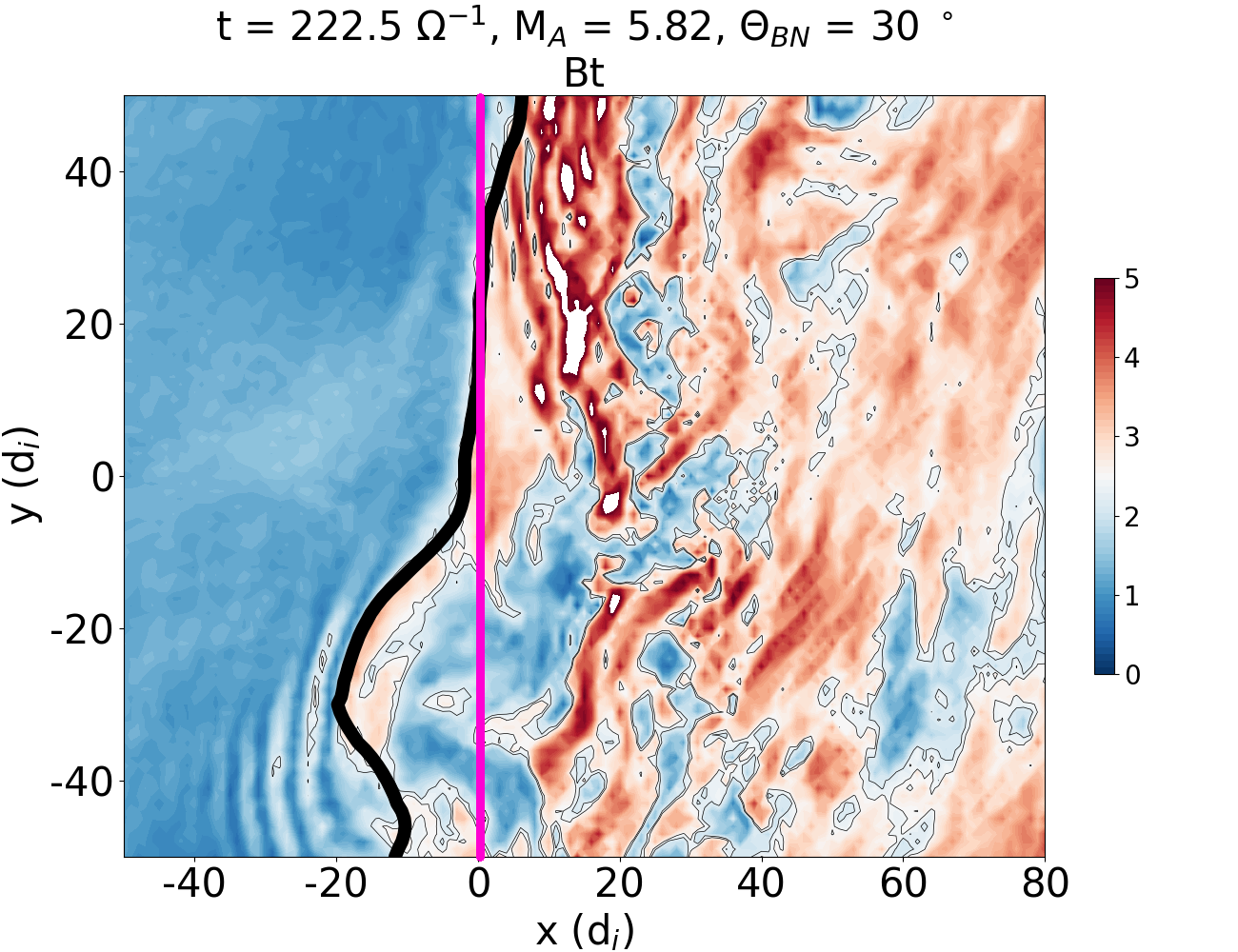}
 \caption{An excerpt from our run {\it e} with the thick black curve representing the shock surface and the vertical purple line marking the average shock location. Here we use a coordinate system that is shifted along $x$, so that the average shock position is at $x$ = 0.}
 \label{fig:bf089}
\end{figure}

In the Section~\ref{sec:theory} we showed that the wavelength of the upstream ULF waves depends on M$_A$ and $\theta_{BN}$ and that the dependence on the latter is much stronger. We also showed that the amplitude of these waves depends on the product of the reflected beam density and velocity, both of which also depend on the same two shock parameters. Hence, the ULF wave properties are a complex function of M$_A$ and $\theta_{BN}$, as can be seen in Figures~\ref{fig:bz} and \ref{fig:bf}. If ripple properties depend on the properties of the upstream waves, we may expect similarly complex dependence of the degree of rippling on M$_A$ and $\theta_{BN}$.

\begin{landscape}
\begin{figure}[h]
\centering
\includegraphics[height=1.0\textheight]{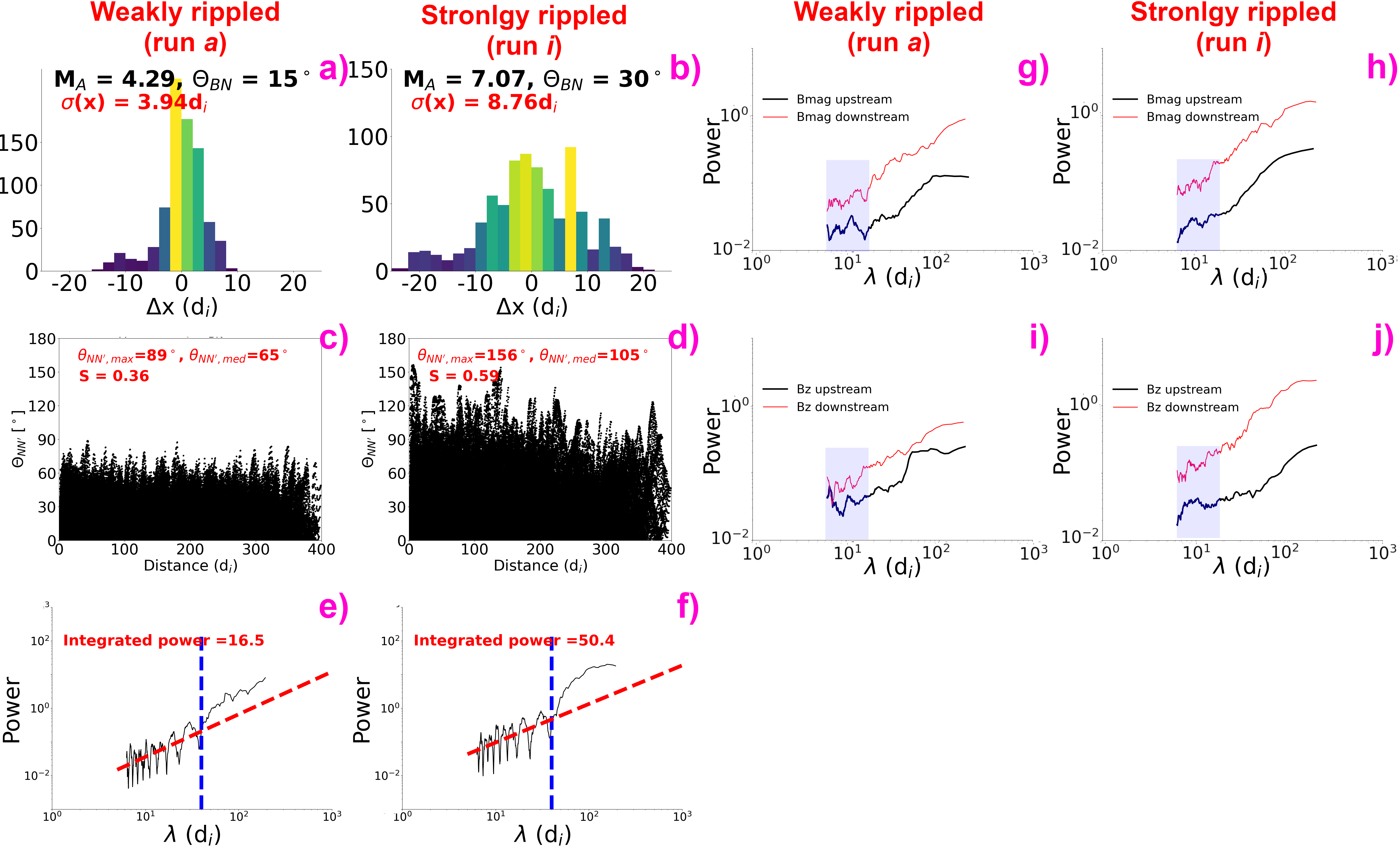}
 \caption{a), b) Histograms of distances $\Delta x$ of the points on the shock surface from the shock's mean location. c), d) $\theta_{NN'}$ angles for each pair of normals calculated for all points on the shock surfaces. e), f) Power spectra of the curves representing the shock surfaces. Left panels represent data for the weakly rippled shock from the run {\it a}, while right panels are for a strongly rippled shock from run {\it i}.}
 \label{fig:figure4}
 \end{figure}
\end{landscape}

\begin{figure}[h]
\centering
\includegraphics[width=1.0\textwidth]{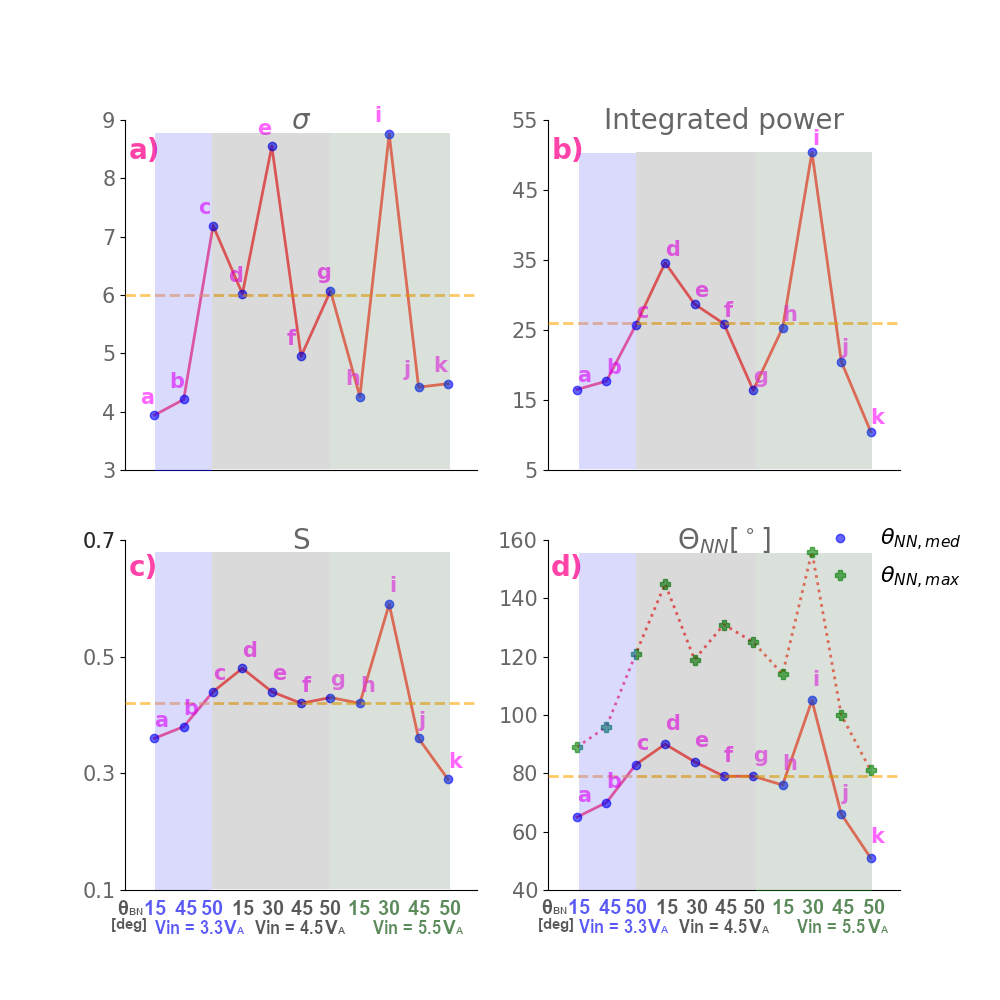}
 \caption{Values of a) $\sigma$, b) inte, c) $S$ and d) maximum and median $\theta_{NN'}$ for all the runs.}
 \label{fig:figure5}
 \end{figure}

\begin{figure}[h]
\centering
\includegraphics[width=1.0\textwidth]{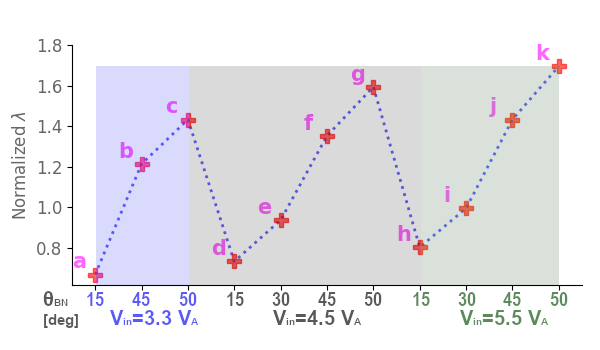}
 \caption{Predicted normalized wavelength ($\lambda/\frac{2\pi V_A}{\Omega_p})$) for different models. $\theta_{BN}$ angles are stated for each model.}
 \label{fig:figure13}
 \end{figure}

We determine how rippled the shocks are by studying how certain characteristics of the ripples (amplitudes, length scales and the steepness of the ripple sides) vary between runs. We characterize the ripple amplitudes with the dispersion $\sigma$ of the distributions of distances of points $\Delta$x on
shock surfaces from the mean shock locations (Figures~\ref{fig:figure4}a, b and \ref{fig:figure5}a). The spatial scales of the ripples are studied with the Fourier spectra of the curves that we use to describe the surfaces (Figures~\ref{fig:figure4}c, d and \ref{fig:figure5}b). Finally, the $\theta_{NN'}$ plots describe the steepness of the sides of the ripples  (Figures~\ref{fig:figure4}e, f and \ref{fig:figure5}c, d). Based on these statistics we divide the shocks into three groups: the weakly rippled shocks with small $\sigma$, mean $\theta_{NN'}$, $S$ and Fourier power and the strongly rippled shocks for which these quantities are larger. The runs {\it c)} and {\it g} are considered intermediately rippled since they exhibit low values of integrated power (Figure~\ref{fig:appendix2}). 

There is a correlation between the upstream waves and the shock ripples. Many short-wavelength waves produce many smaller-scale ripples. These exhibit histograms with small standard deviations ($\sigma$) and many narrow peaks in $\theta_{NN}$ plots (see panels \ref{fig:bz}a, \ref{fig:bf}a, \ref{fig:figure4}a and \ref{fig:figure5}a). On the other hand few long wavelength upstream waves tend to produce less large-scale ripples and result in fewer thicker peaks in $\theta_{NN}$ plots (compare panels \ref{fig:bz}i, \ref{fig:bf}i, \ref{fig:figure4}b and \ref{fig:figure5}a). Large (small) amplitude waves tend produce shock ripples with larger (smaller) amplitudes and histograms with larger $\sigma$.

We also compare properties of upstream and downstream fluctuations. Most upstream spectra exhibit flattening or enhancements in the ULF range. These features are conserved in the spectra of the downstream compressive fluctuations. Flattening is better conserved in the case of weakly rippled shocks. Downstream transverse fluctuations however exhibit spectra that are more turbulence-like with flattening and/or enhancements conserved to a much lesser. This is probably due to the fact that downstream fluctuations are mostly compressive.

\begin{landscape}
\begin{figure}[h]
\centering
\includegraphics[height=0.9\textheight]{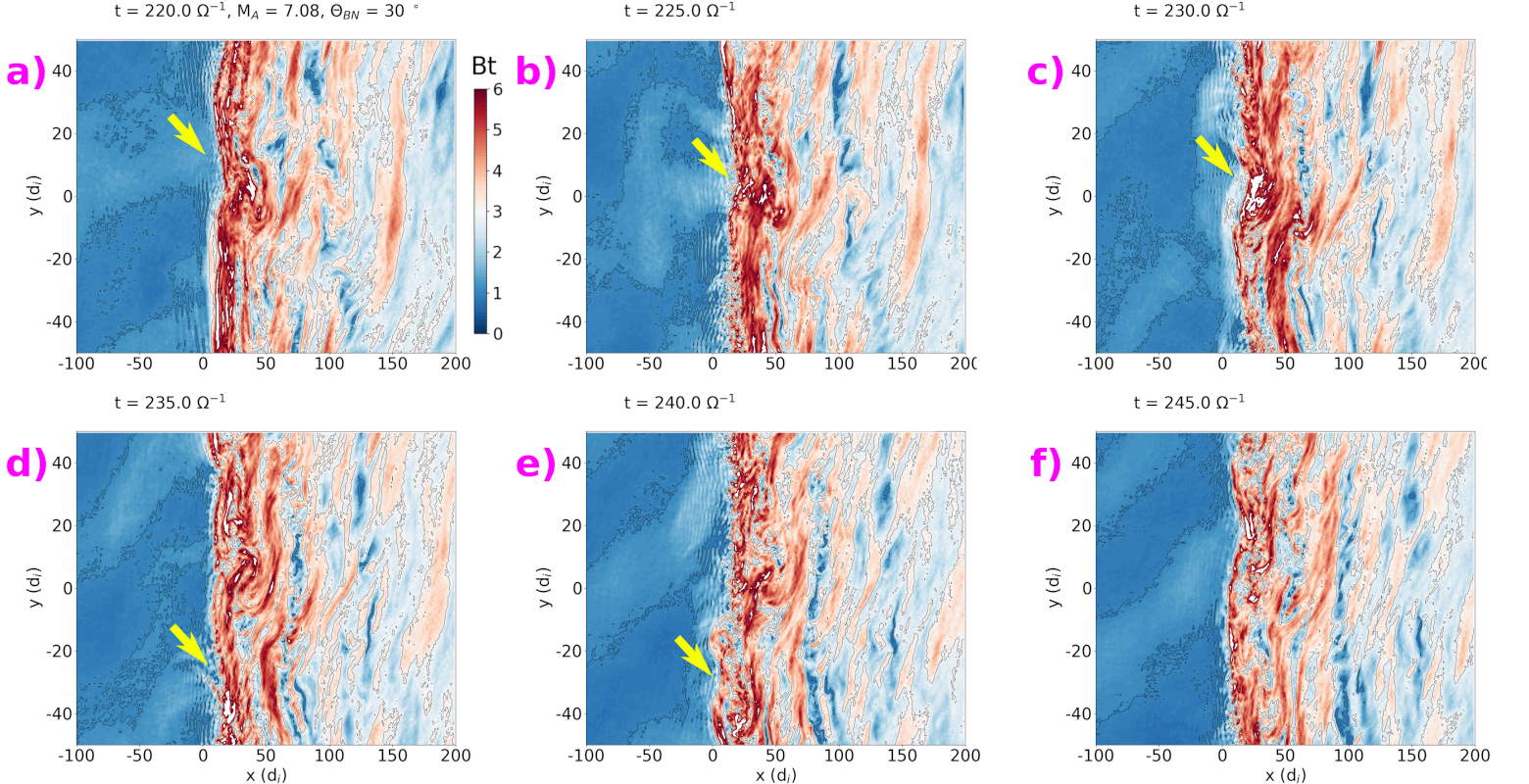}
 \caption{Excerpts from the outputs of our run {\it i} at five different moments. The colors represent the B magnitude.}
 \label{fig:bf30}
 \end{figure}
\end{landscape}

\begin{landscape}
\begin{figure}[h]
\centering
\includegraphics[height=0.9\textheight]{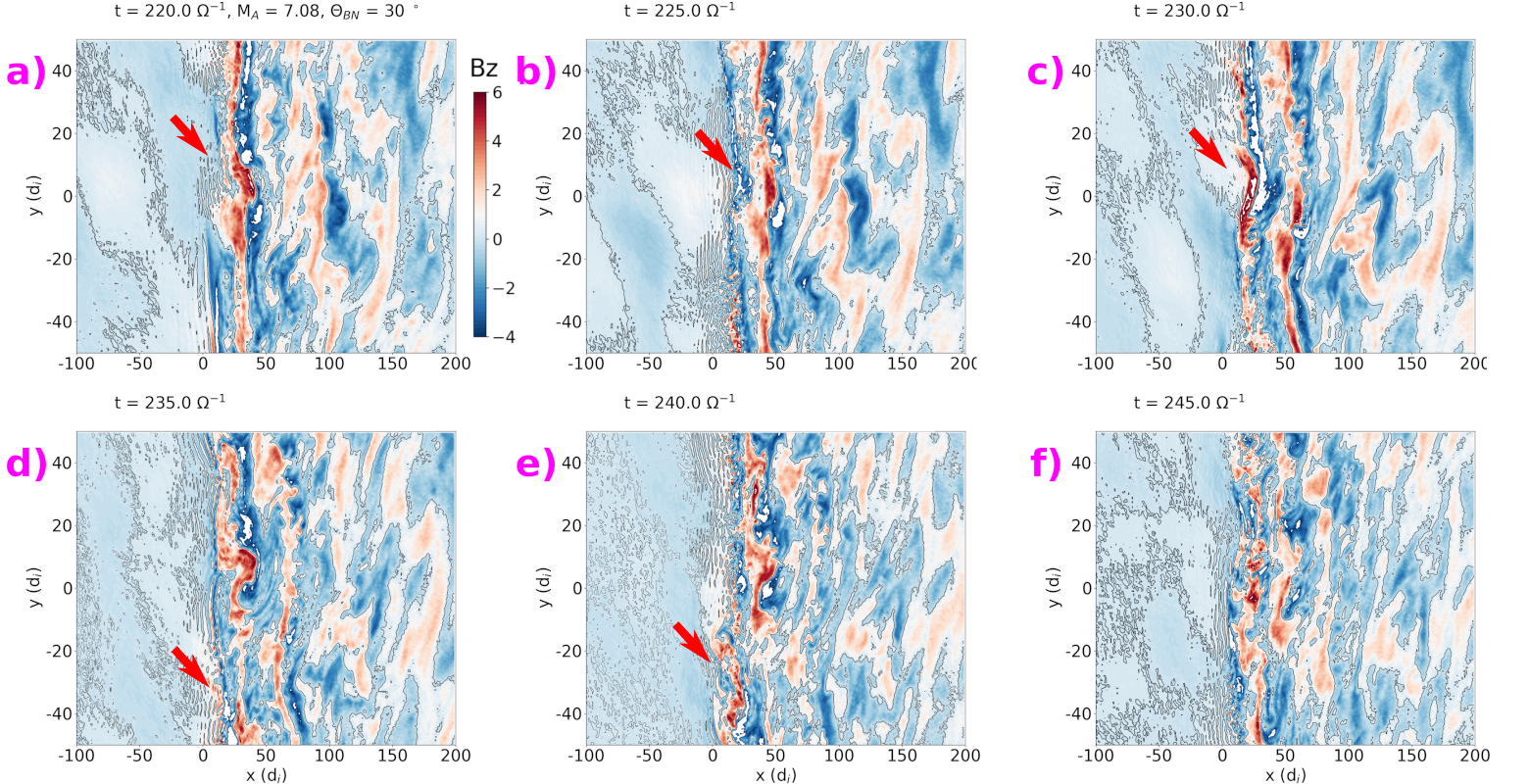}
 \caption{Excerpts from the outputs of our run {\it i} at five different moments. The colors represent the B$_z$ component.}
 \label{fig:bz30}
 \end{figure}
\end{landscape}

By examining a case study (Figures~\ref{fig:bf30} and \ref{fig:bz30}) we see that it is the compressive waves that locally bend the shock surface, creating concave ripples. However since compressive fluctuations are not aligned with the shock surface, different portions of these waves interact with different sections of the shock at different times. As a consequence the ripples follow this interaction region and thus travel along the shock surface. This also means that different portions of the waves encounter the shock with different properties (magnetic field jump, local normal orientation, etc.). This results in the loss of the wave's identity. Downstream fluctuations do not resemble the upstream waves in wavelengths, wavefront extension and orientation nor amplitude. This is the reason why the transverse downstream Fourier spectra Figures~\ref{fig:figure4}i, j and \ref{fig:appendix4} are much more similar to turbulent spectra. Even in the case of the compressive downstream spectra Figures~\ref{fig:figure4}g, h and \ref{fig:appendix5}, the features they exhibit are less prominent than in the case of upstream spectra.

The fact that flattening and bumps in the ULF range are conserved may be important in the context of ground magnetic pulsations, such as those in Pc2, Pc3 and Pc4 (100-200 mHz, 22-100 mHz and 7-22 mHz, respectively) bands. Their appearance is known to be associated with the IMF orientation and their origin is commonly attributed to the foreshock ULF waves since both exhibit similar frequencies \citep[e.g. ][]{ engebretson:1987, vero:1998,clausen:2009}. The present work shows that some spectral features of compressive fluctuations in the ULF range may indeed survive in the magnetosheath and perhaps affect the magnetopause where they could act as pressure pulses exciting magnetospheric Pc fluctuations. Although local runs provide valuable information on the nature of the interaction between the shocks and the upstream ULF waves, further investigation with global hybrid runs is required in order to study the transmission of such fluctuations from the bow-shock all the way to the magnetopause.

\begin{table}
\centering
\begin{tabular}{l c c c}
 \hline
  V$_{in}$  & $\theta_{BN}$ & $\beta$ & M$_A$ \\
  V$_A$ & $^\circ$ & &\\
 \hline
3.3 & 15, 45, 50 & 0.5 & 4.29, 4.48, 4.57\\
4.5 & 15, 30, 45, 50 & 0.5 & 5.31, 5.82, 5.94, 6.05\\
5.5 & 15, 30, 45, 50 & 0.5 & 6.97, 7.07, 7.17, 7.42\\
\hline
\end{tabular}
\caption{Initial conditions of 2.5D HYPSI runs.}
\label{tab:hypsi}
\end{table}

\noindent{\bf Acknowledgments}\\
We acknowledge the International Space Science Institute for supporting the ISSI team 448 on ``Global study of the transmission of foreshock ULF waves into the magnetosheath and the magnetosphere'' led by L. Turc and M. Palmroth, in which this study was initiated. The simulation outputs from which the Figures~\ref{fig:bf} and \ref{fig:bz} were created are available at https://zenodo.org/record/5156128. PK's work was supported by DGAPA/PAPIIT grant IN105620. The work of N.T. was supported by Grants-in-Aid for Scientific Research of Japan Society for the Promotion of Science (grant number: 16H06286, 18KK0099). K.T.'s work was made supported by the NASA Grant NNX17AD34G. The work of L.T. is supported by the Academy of Finland (grant number 322544). EK acknowledges the ERC under the European Union's Horizon 2020 Research and Innovation Programme Project SolMAG 724391, and Academy of Finland Project 310445." X.B.C.’s work was supported by UNAM DGAPA PAPIIT IN-105218-3 grant. APD received financial support from the Swedish National Space Agency (Grant \#2020-00111) and the EU Horizon 2020 project SHARP – SHocks: structure, AcceleRation, dissiPation \# 101004131. Y.P.K.'s work was supported by the ERC Consolidator grant no. 682068-PRESTISSIMO.

\section{Appendix}

\begin{landscape}
\begin{figure}[h]
\centering
\includegraphics[height=1.0\textheight]{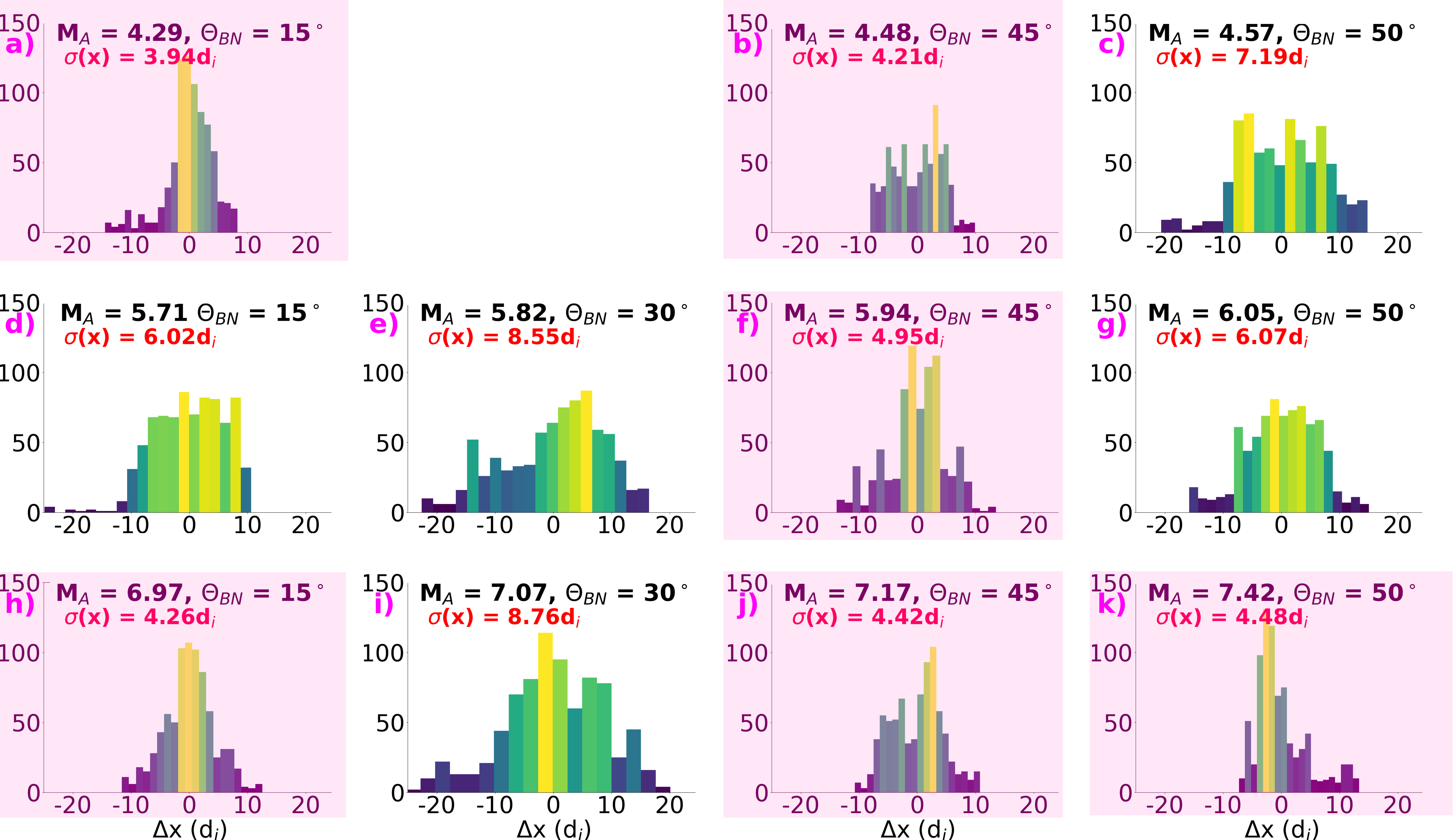}
 \caption{Histograms for distributions of x-coordinates along the shock surface with respect to its average position. Shaded panels represent what we classify as weakly rippled shocks.}
 \label{fig:appendix1}
 \end{figure}
\end{landscape}

\begin{landscape}
\begin{figure}[h]
\centering
\includegraphics[height=1.0\textheight]{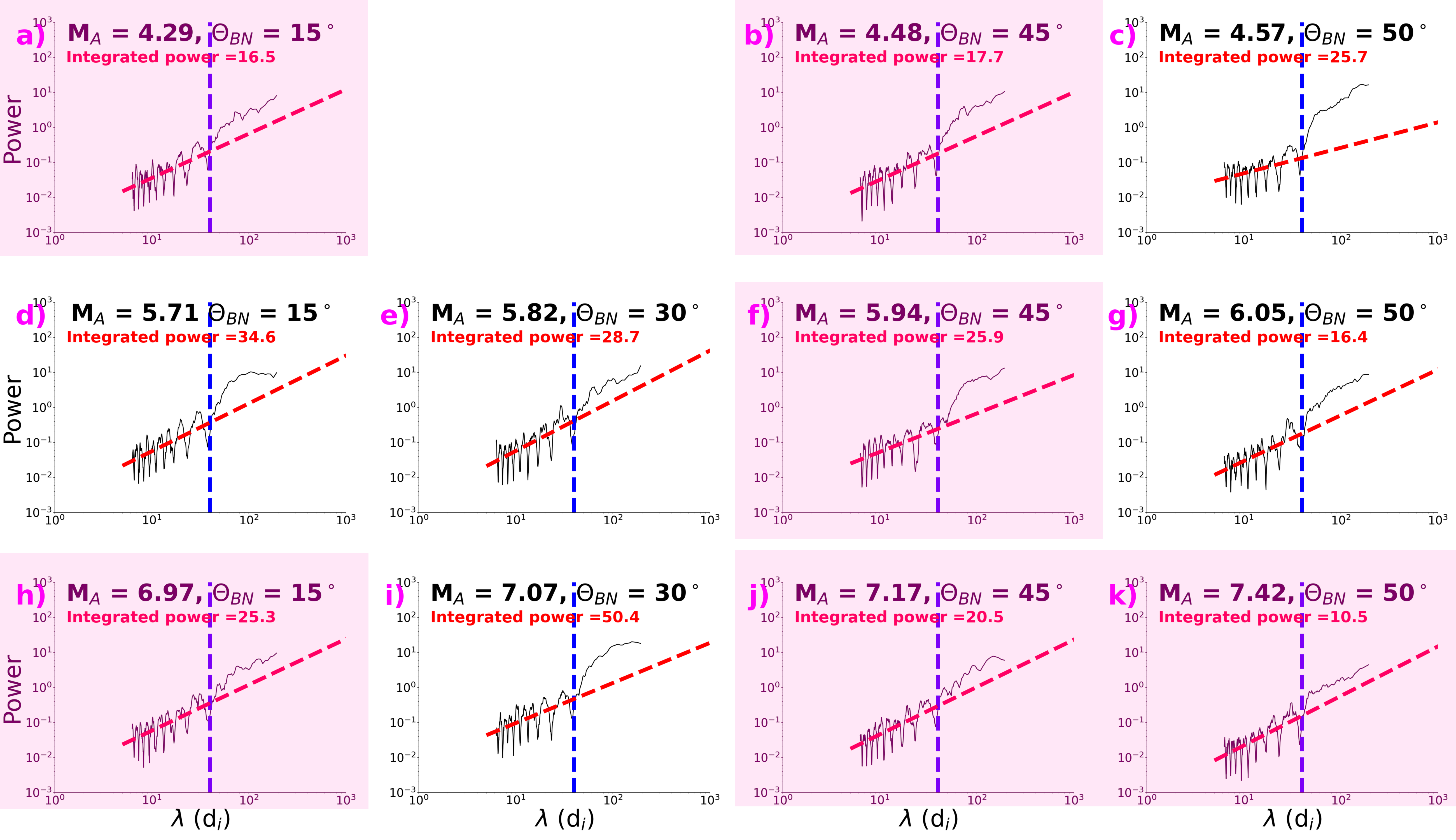}
 \caption{Fourier spectra of curves representing the shock surfaces. Vertical blue line marks the location of $\lambda$=40~d$)_i$, while the red line marks the trends at $\lambda<$40~d$)_i$. The integarted power is calculated by calculating the total power of the spectra at $\lambda >$40~d$_i$ and subtracting from it the power delimited by the red line in the same wavelength range. Shaded panels represent what we classify as weakly rippled shocks.}
 \label{fig:appendix2}
 \end{figure}
\end{landscape}

\begin{landscape}
\begin{figure}[h]
\centering
\includegraphics[height=1.0\textheight]{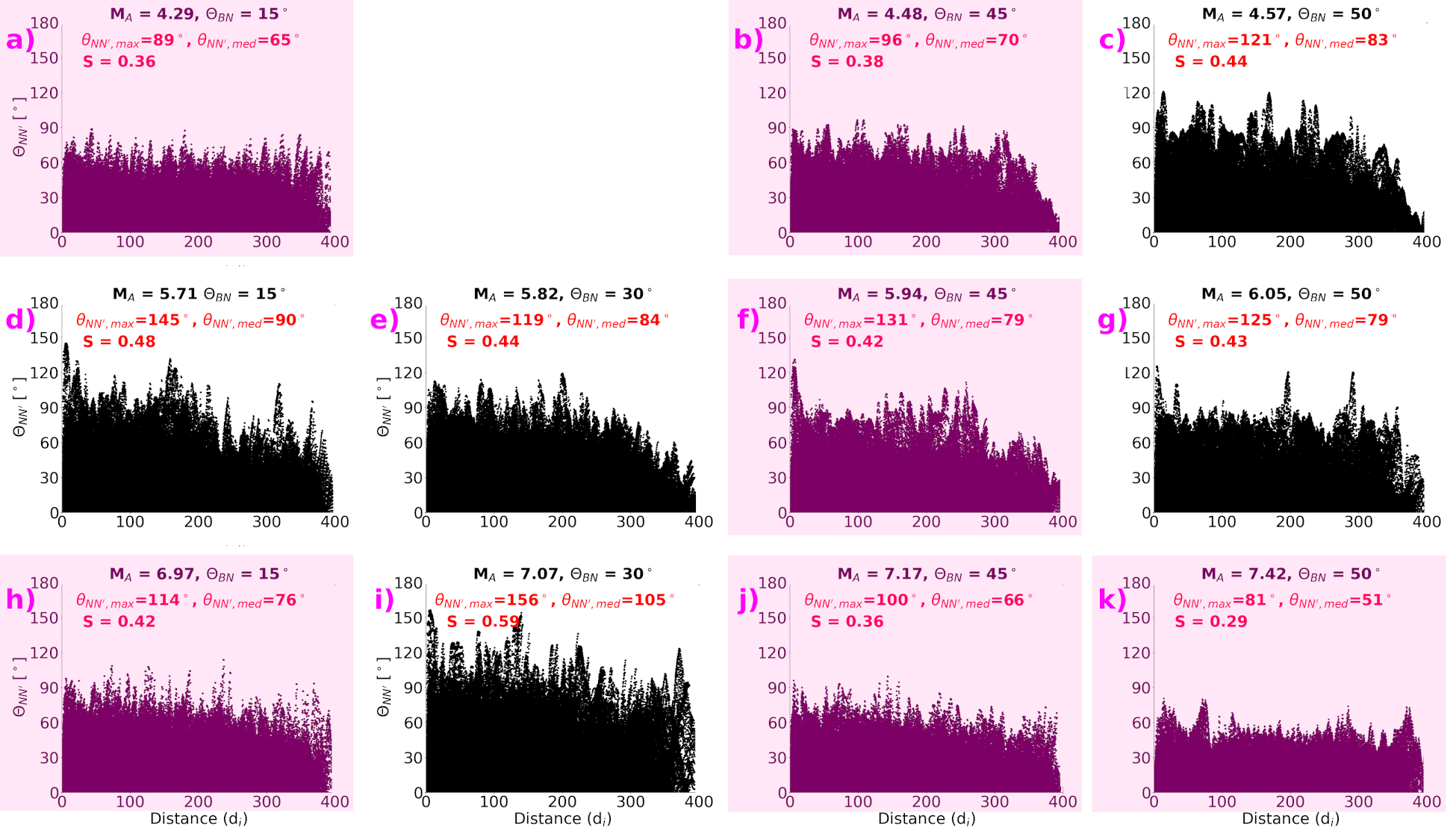}
 \caption{$\theta_{NN}$ angles between the normals of all pairs of points along the shock surface as a function of distance between the points. The maximum and median $\theta_{NN'}$ and the value S, which is the fraction of the surface delimited by black dosts, are also shown. Shaded panels represent what we classify as weakly rippled shocks.}
 \label{fig:appendix3}
 \end{figure}
\end{landscape}

\begin{landscape}
\begin{figure}[h]
\centering
\includegraphics[height=1.0\textheight]{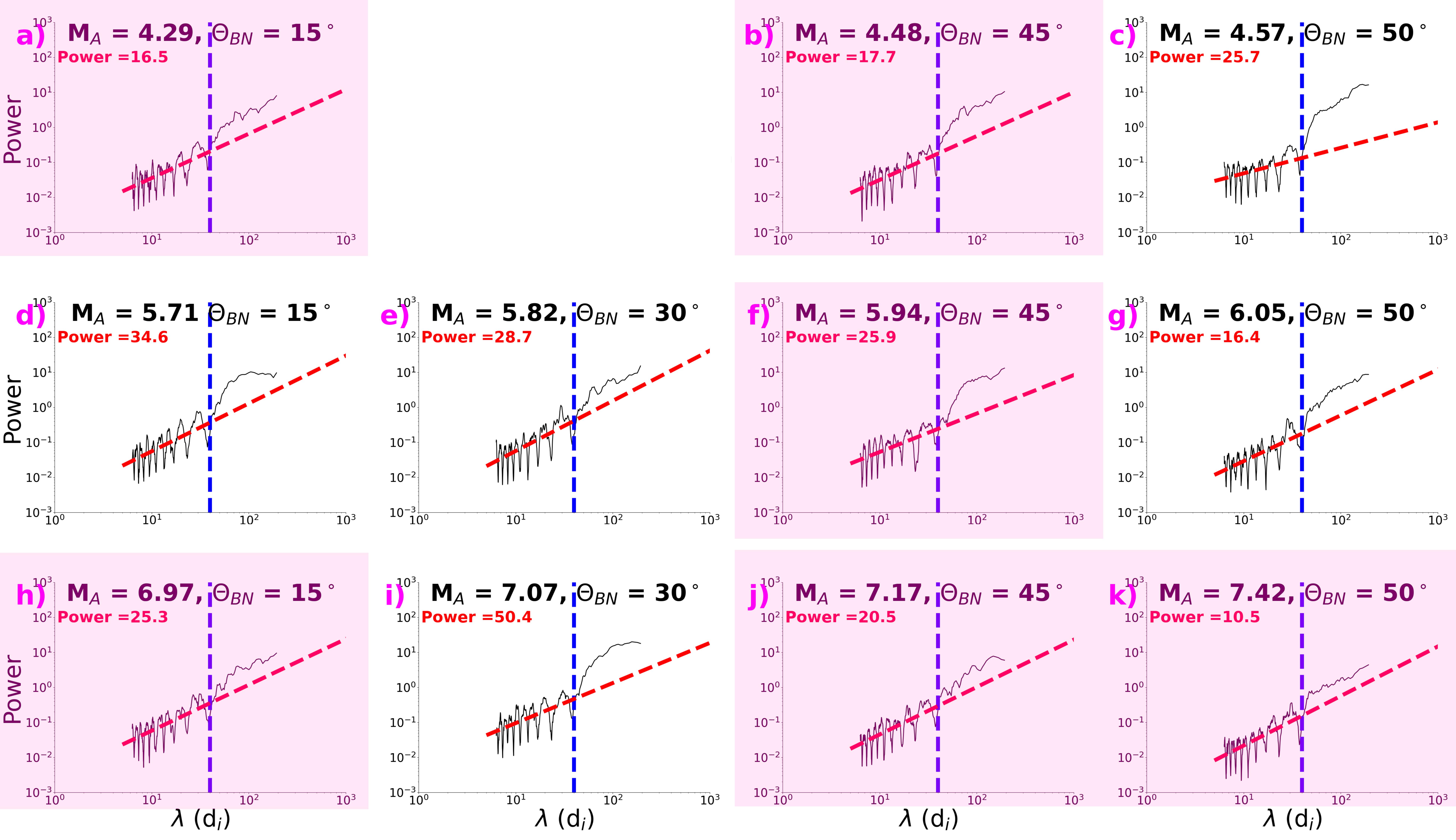}
 \caption{Fourier spectra of curves representing the shock surfaces. Vertical blue line marks the location of $\lambda$=40~d$_i$, while the red line marks the trends at
$\lambda<$40~d$_i$. The integrated power is calculated by calculating the total power of the spectra at $\lambda>$40~d$_i$ and subtracting from it the power delimited by the red line in
the same wavelength range. Shaded panels represent what we classify as weakly rippled shocks.}
 \label{fig:appendix4}
 \end{figure}
\end{landscape}

\begin{landscape}
\begin{figure}[h]
\centering
\includegraphics[height=1.0\textheight]{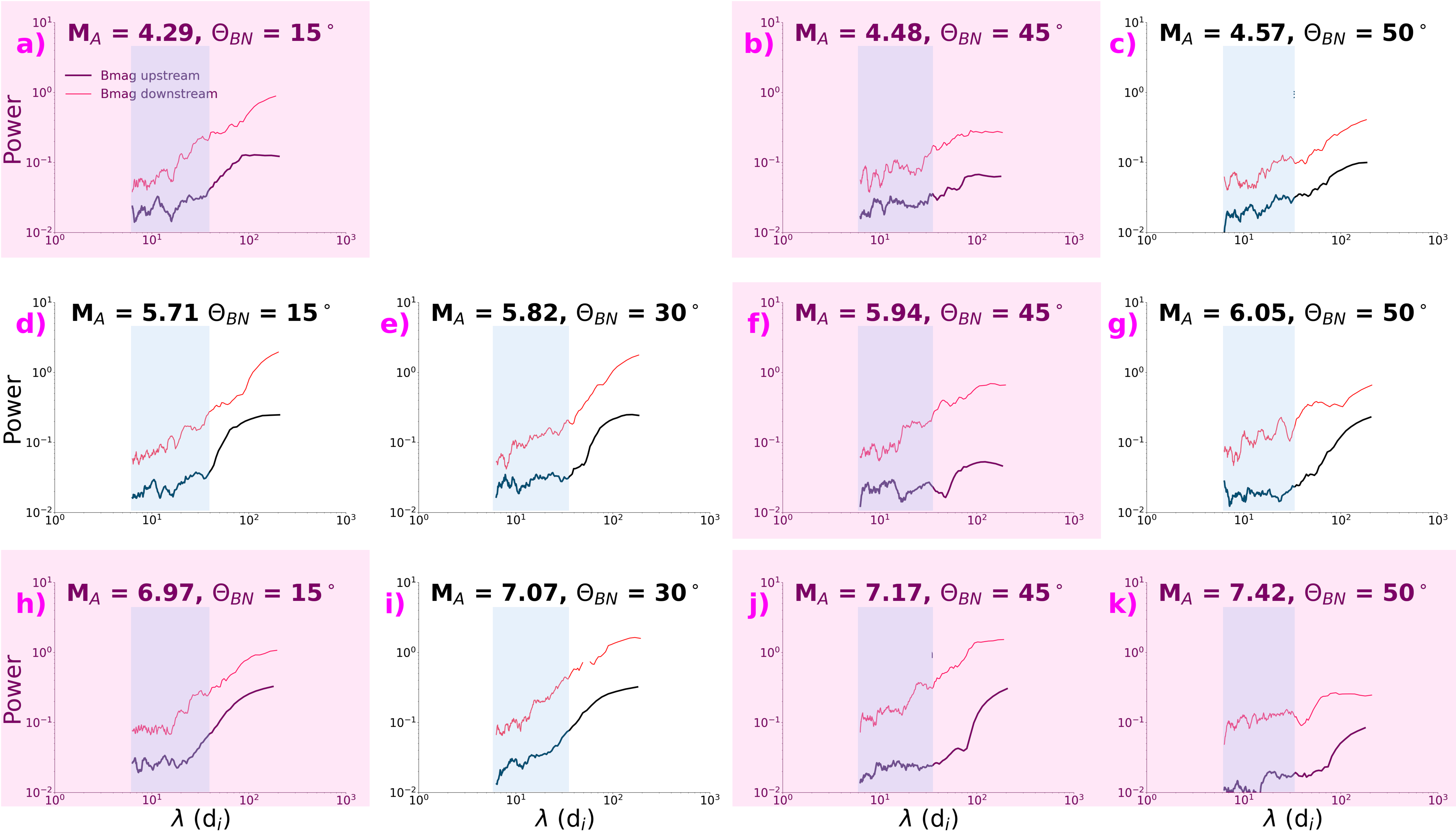}
 \caption{Upstream (black) and downstream (red) Fourier spectra of $|$B$|$ fluctuations at times shown in Figure~\ref{fig:bf}. Blue shading indicates the ULF range. Panels shaded in pink represent what we classify as weakly rippled shocks.}
 \label{fig:appendix5}
 \end{figure}
\end{landscape}

\begin{landscape}
\begin{figure}[h]
\centering
\includegraphics[height=1.0\textheight]{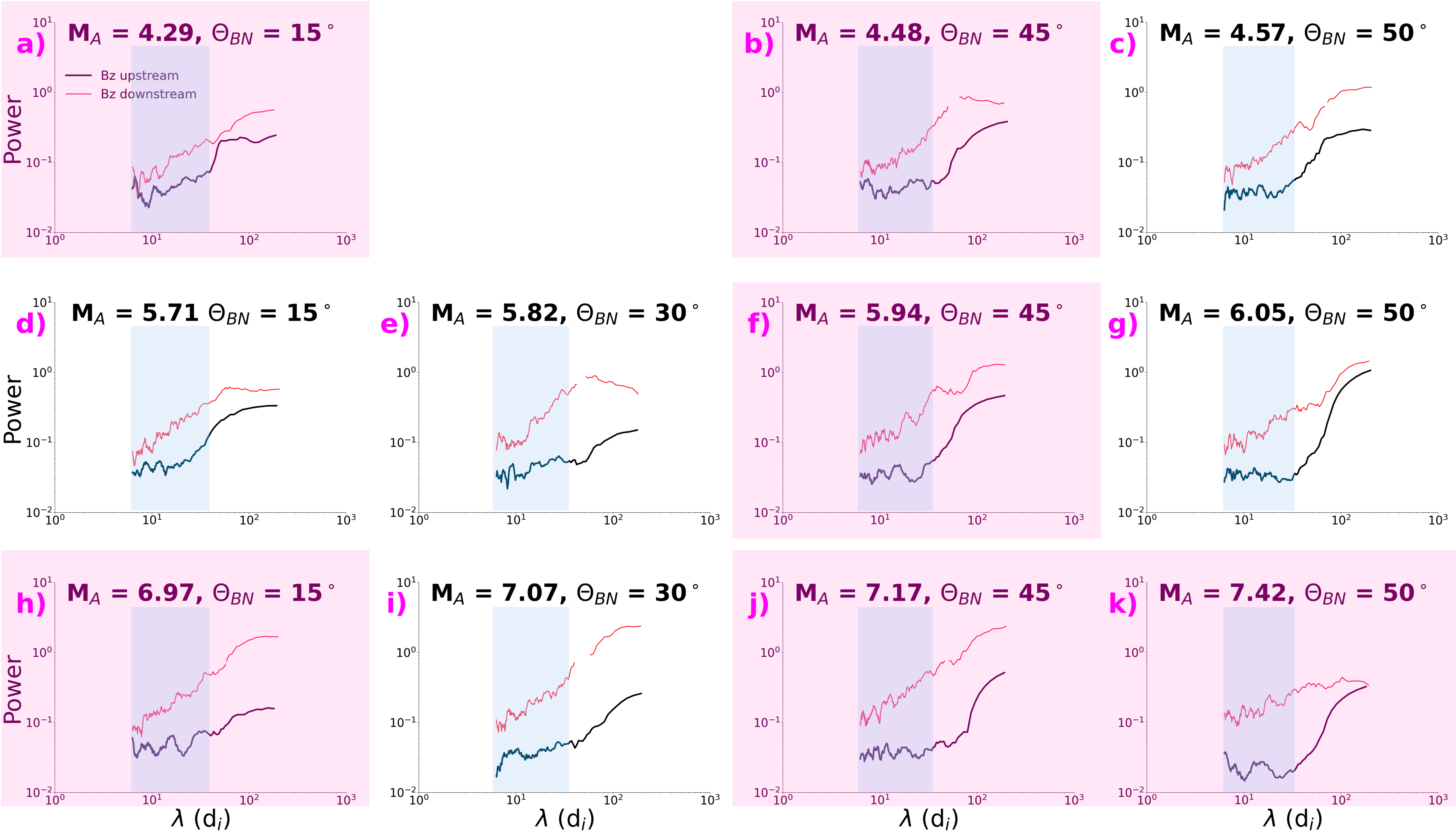}
 \caption{Upstream (black) and downstream (red) Fourier spectra of B$_z$ fluctuations at times shown in Figure~\ref{fig:bz}.  Blue shading indicates the ULF range. Panels shaded in pink represent what we classify as weakly rippled shocks.}
 \label{fig:appendix6}
 \end{figure}
\end{landscape}



\end{document}